\documentclass[journal]{IEEEtran}
\makeatletter
\def\subsubsection{
  \@startsection
    {subsubsection}
    {3}
    {\parindent}
    {3.5ex plus 1.5ex minus 1.5ex}
    {0.7ex plus .5ex minus 0ex}     
    {\normalfont\normalsize\itshape}
}
\makeatother
\usepackage{amsmath,amsfonts}
\usepackage{algorithmic}
\usepackage{algorithm}
\usepackage{array}
\usepackage[caption=false,font=footnotesize]{subfig}
\usepackage{textcomp}
\usepackage{stfloats}
\usepackage{url}
\usepackage{verbatim}
\usepackage{graphicx}
\usepackage{cite}
\usepackage{hyperref}
\usepackage{url}
\usepackage{cleveref}
\usepackage{siunitx}
\usepackage[export]{adjustbox}

\begin{document}

\title{Power Reserve Capacity from Virtual Power Plants with Reliability and Cost Guarantees}

\author{Lorenzo~Zapparoli,
        Blazhe~Gjorgiev,
        Giovanni~Sansavini,~\IEEEmembership{Member,~IEEE}
\thanks{Lorenzo Zapparoli, Blazhe Gjorgiev, and Giovanni Sansavini are with the Institute of Energy and Process Engineering, ETH Zurich, Zurich, Switzerland (e-mail: lzapparoli@ethz.ch, gblazhe@ethz.ch, sansavig@ethz.ch)}}

\maketitle
\begin{abstract}
The growing penetration of renewable energy sources is expected to drive higher demand for power reserve ancillary services (AS). One solution is to increase the supply by integrating distributed energy resources (DERs) into the AS market through virtual power plants (VPPs).
Several methods have been developed to assess the potential of VPPs to provide services. However, the existing approaches fail to account for AS products' requirements (reliability and technical specifications) and to provide accurate cost estimations.
Here, we propose a new method to assess VPPs’ potential to deliver power reserve capacity products under forecasting uncertainty. First, the maximum feasible reserve quantity is determined using a novel formulation of subset simulation for efficient uncertainty quantification. Second, the supply curve is characterized by considering explicit and opportunity costs.
The method is applied to a VPP based on a representative Swiss low-voltage network with a diversified DER portfolio. We find that VPPs can reliably offer reserve products and that opportunity costs drive product pricing.
Additionally, we show that the product's requirements strongly impact the reserve capacity provision capability.
This approach aims to support VPP managers in developing market strategies and policymakers in designing DER-focused AS products.
\end{abstract}


\section{Introduction}
\label{intro}

\IEEEPARstart{T}{he} increasing penetration of renewable energy sources (RES) and distributed energy resources (DERs) is radically transforming the power system, introducing significant challenges. Primarily, renewable energy generation's inherent variability is expected to increase the demand for power reserves to guarantee secure grid operation. This would force transmission system operators (TSO) to procure larger reserve capacities on the ancillary services (AS) markets. Such a trend coincides with a decline of traditional AS providers, potentially leading to AS shortages~\cite{xie2024}. One solution to this challenge is integrating DERs into AS markets to enhance liquidity and reduce costs by increasing competition. Virtual power plants (VPP) address this by pooling numerous small-scale DERs to efficiently trade energy and provide grid support services as a single, reliable resource~\cite{pudjianto2007}. 

Power reserve capacity products are subject to strict technical requirements regarding duration, ramp time, and reliability. In particular, the reliability requirement, which typically exceeds \(\SI{99}{\%}\)~\cite{camal2023}, poses a challenge for providing power reserve capacity with DERs, as reserves are booked ahead of delivery when forecasting uncertainty is relevant. To participate in the reserve capacity AS markets, aggregators must be able to adequately estimate extreme quantiles of the expected AS product availability distribution (\(\SI{0.1}{\%}\) to \(\SI{1}{\%}\)). This requires the consideration of the various factors that may jeopardize reserve availability in real-time operation, such as the DERs' technical limitations, network limits, and forecast uncertainties. In addition, the cost of the reserve product needs to be addressed~\cite{silva2018, wang2024}.

Existing literature predominantly considers technical VPPs, which are linked to an active distribution network (ADN) and incorporate grid constraints in their operations schemes~\cite{pudjianto2007, tan2020}. The flexibility of a VPP is typically characterized by the feasible operating region (FOR)~\cite{churkin2024}, also referred to as the P-Q area. The FOR includes the set of feasible power exchanges between the VPP and the upstream network at their point of common coupling (PCC) in the P-Q power plane~\cite{Contreras2021}. 
The FOR is typically determined through an optimization-based approach. First, a problem is formulated to maximize the power exchange at the PCC along a specific direction on the P-Q plane subject to DERs and network constraints. Then, the FOR contour is characterized by solving the optimization problem multiple times along different directions, e.g. using an angle-based contour search algorithm~\cite{silva2018}.
In~\cite{Riaz2022}, the authors include some technical requirements of power reserve products in the optimization problem. The paper highlights their significant impact on the FOR and stresses the need to account for forecasting uncertainty in the flexibility assessment.

Neglecting uncertainty can lead to overestimating the available reserve, non-compliance with reliability requirements, and the inability to account for cost uncertainty in product pricing. Uncertainty has been addressed in literature via 1) Monte Carlo simulation \cite{gonzales2018}, 2) robust optimization \cite{kalantarneyestanaki2020}, or 3) chance constraints \cite{wang2024}. The authors in~\cite{gonzales2018} consider forecasting uncertainty through a Monte Carlo simulation. However, the computational time required by the full AC power flow model limits the simulations to very few samples. The network model is linearized in~\cite{kalantarneyestanaki2020}, and scenario-based robust optimization is adopted. 
The robust approach allowed for a drastic reduction in the computation time but led to overconservative solutions~\cite{maMmaRELLA2022110108}.
In~\cite{wang2024}, the authors adopt chance constraints, achieving good accuracy with low computational effort.
However,~\cite{zhou2024} observes that individual chance constraints are often insufficient to guarantee the target reliability. Instead, distributionally robust joint chance constraints (DRJCC) are proposed. To make the problem with DRJCC tractable, a surrogate polytope is used to approximate the FOR. The authors show how this method can meet the reliability requirements but may be overconservative and not always computationally tractable. In conclusion, the proposed uncertainty modeling techniques are either over/under-conservative or too computationally expensive.

Besides the lack of appropriate approaches to account for forecasting uncertainty, the use of the FOR as a proxy for reserve capacity should be challenged. Although the FOR promises to be an effective flexibility metric, it does not reflect how power reserves are currently booked. TSOs rely on standardized power reserve capacity products with specific technical requirements. Quantifying the amount of such products that a VPP can provide and determining the associated costs is critical for integrating larger shares of DERs into existing market structures. This aspect has yet to receive significant attention in the literature.

We address these gaps by proposing a novel method for assessing the power reserve provision capability of a VPP under uncertainty. This method characterizes the supply curve of reserve capacity products, allowing the quantification of products' maximum quantity and costs.

The method has two steps. First, we determine the maximum product quantity the VPP can provide. Here, uncertainty is considered using a stochastic approach, introducing a novel subset-simulation-based quantile estimator to model the reliability requirement with reduced computational effort. In the second step, the product cost is characterized as a function of quantity, considering both explicit DER costs and opportunity costs. The method is applied to a VPP based on a realistic Swiss low-voltage network, considering a comprehensive set of flexible resources, namely distributed generators (DG), electric vehicles (EV), heat pumps (HP), and battery energy storage systems (BESS).

The contributions of this paper are: 1) a method for assessing the provision of power reserve capacity products by a VPP under uncertainty, via the characterization of the product supply curve considering network, DERs, and product constraints; 2) an efficient subset-simulation-based quantile estimator for the assessment of the product provision reliability; 3) an analyses on how the technical product requirements (reliability, ramp time, and duration) affect the VPP’s ability to offer power reserves for a realistic case study.
The proposed method can be used by VPP managers to design bidding strategies, by network planners to estimate the reserves that can be gathered from distribution networks, and by policymakers to design new ancillary service products tailored for DERs.

The remainder of this paper is organized as follows. Section~\ref{sec:method} describes the proposed method. Section~\ref{sec:cstudy} introduces the case study used to demonstrate the method. Section~\ref{sec:results} provides results followed by a discussion thereof. Section~\ref{sec:conclusions} concludes and provides an outlook for future work.

\section{Method}
\label{sec:method}

\begin{figure}
    \centering
    \includegraphics[width=1.00\linewidth]{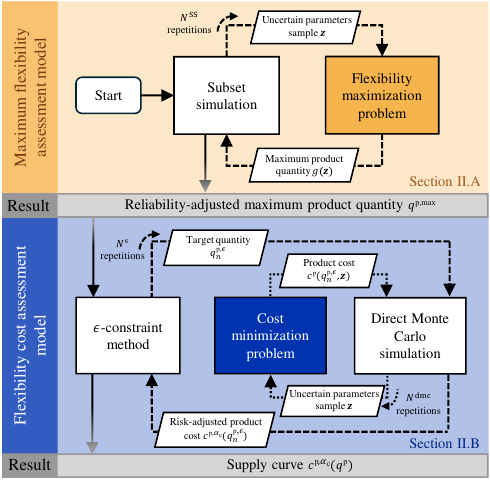}
    \caption{High-level architecture of the two-step method for determining the supply curve of power reserve capacity products. Rectangles represent algorithms/models, parallelograms represent data, and dashed lines indicate repeated actions, with the number of executions specified by corresponding dashed arrows.}
    \label{fig:high_level_method_architecture}
\end{figure}

This section presents the two-step approach schematically represented in Fig.~\ref{fig:high_level_method_architecture}.
In the first step, the maximum flexibility (i.e., product quantity) the VPP can provide is determined through the maximum flexibility assessment model. In the second step, the flexibility cost assessment model characterizes the cost of the power reserve products as a function of quantity.

\subsection{Maximum flexibility assessment model}
\label{sec: method maximum flexibility assessment model}

Reserve capacity products are procured on the AS market in a tendering procedure that takes place at time \(t^{\text{lead}}\), ahead of the delivery time \(t^{\text{d}}\). Each product is characterized a direction \(d^{\text{p}} \in \{\text{upward, downward, symmetrical}\}\), and covers a specific delivery period \(T^{\text{d}} = [t^{\text{d}}, t^{\text{d}} + t^{\text{p}}]\), where \(t^{\text{p}}\) is the product duration. The booked capacity \(q^{\text{p}}\) must be available throughout \(T^{\text{d}}\) to accommodate any reserve activation request by the TSO up to \(q^{\text{p}}\), with a maximum ramp time \(r^{\text{p}}\) and a reliability \(R^{\text{p}}\)~\cite{swissgrid2022_ASproducts, swissgrid2022_prequalification}.

The model developed here determines the maximum reserve capacity product quantity a VPP can offer without violating network and DER constraints while respecting the product's technical requirements. 
To account for uncertainty in DER forecasts at \(t^{\text{lead}}\), the flexibility maximization problem determines the maximum reserve capacity \(g(\boldsymbol{z})\) the VPP can provide given a realization of the uncertain parameters \(\boldsymbol{z} \in \mathcal{P}^{\text{u},\text{mfa}}\), described in Section \ref{sec: method flexibility maximization model}. Then, the subset simulation procedure propagates uncertainty through \(g(\boldsymbol{z})\) and enforces the reliability requirement, estimating the reliability-adjusted maximum product quantity \(q^{\text{p,max}}\).

The flexibility maximization problem is an optimization problem that finds the network's and DERs' operational state that maximizes the reserve capacity availability \(q^{\text{p}}(\boldsymbol{x},\boldsymbol{y},\boldsymbol{z})\). It is formulated as a mixed-integer linear program (MILP):

\begin{equation}
    \begin{aligned}
    g(\boldsymbol{z}) = \ & \underset{\boldsymbol{x},\boldsymbol{y}}{\text{max}} & & q^{\text{p}}(\boldsymbol{x},\boldsymbol{y},\boldsymbol{z}) \\
    & \text{s.t.} & & \text{Product-defining constraints} \\
    & & & \text{Network constraints} \\
    & & & \text{DERs' constraints}
    \end{aligned}
    \label{eq: method flexibility maximization model}
\end{equation}
where \(\boldsymbol{x}\) and \(\boldsymbol{y}\) are continuous and binary decision variables, respectively. Three sets of constraints are applied. The product-defining constraints map electrical quantities to the corresponding product quantity and enforce the product's duration requirement. 
The network constraints model the VPP distribution network and enforce voltage and flow limits.
The DER constraints model the operational limits of the flexible resources in the VPP and model the product ramp time requirement.

To satisfy the reliability requirement \(R^{\text{p}}\), parametric uncertainty on \(\boldsymbol{z}\) must be propagated through the flexibility maximization problem \(g(\boldsymbol{z})\), and the \(\alpha^{\text{p}} = 1 - R^{\text{p}}\) quantile of the maximum product quantity must be detemined. Therefore, the reliability-adjusted maximum reserve capacity product quantity \(q^{\text{p,max}}\) is defined as:

\begin{equation}
q^{\text{p,max}} = \inf \left\{ \tilde{q^{\text{p}}} \in \mathbb{R} \mid \mathbb{P} \left(g(\boldsymbol{z}) \leq \tilde{q^{\text{p}}} \right) = \alpha^{\text{p}} \right\}.
\label{eq: quantile_reliability_constraint_definition}
\end{equation}
High-reliability levels (e.g., \(\SI{99}{\%}\)-\(\SI{99.9}{\%}\)) entail an extreme quantile estimation problem. Estimating extreme quantiles using the direct Monte Carlo (DMC) method requires a large number of evaluations, making it computationally expensive~\cite{guyader2011simulation}. In this work, we extend the subset simulation (SS) method, traditionally used in reliability engineering for rare-event probability estimation~\cite{zuev2013subset, doi:https://doi.org/10.1002/9781118398050.ch5}, to estimate extreme quantiles efficiently.
The SS methodology decomposes the original quantile estimation problem into a sequence of simpler conditional quantile estimations, significantly reducing the computational burden.

Section~\ref{sec: method flexibility maximization model} details the flexibility maximization problem, and Section~\ref{sec: method subset simulation} describes the SS method for extreme quantile estimation.

\subsubsection{Flexibility maximization problem}
\label{sec: method flexibility maximization model}
The following describes the formulation of the constraints of the flexibility maximization problem, introduced in \eqref{eq: method flexibility maximization model}, and the uncertain parameters' set \(\mathcal{P}^{\text{u},\text{mfa}}\).

\textit{Product defining constraints.} The VPP is simulated in three states: the dispatching state (disp), which reflects the position acquired in the energy markets; the activated upward reserve state (up), which represents the scenario where the TSO fully utilizes the booked upward reserve; and the activated downward reserve state (down), which represents the scenario where the TSO fully utilizes the booked downward reserve.
The set \( \mathcal{S} \) summarizes the simulated states. It holds that for symmetrical products \( \mathcal{S} = \{ \text{disp}, \text{up}, \text{down} \} \), for upward products \( \mathcal{S} = \{ \text{disp}, \text{up} \} \), and for downward products \( \mathcal{S} = \{ \text{disp}, \text{down} \} \). Network and DERs' constraints are enforced in every state \(s \in \mathcal{S}\) to guarantee the availability of the booked reserve.
For simulation purposes, the delivery period \(T^{\text{d}}\) is discretized using a set of time steps \(\mathcal{T} = \{ t_1, t_2, \dots, t_{T^{\text{d}}/\Delta t} \}\), where each time step has a duration \(\Delta t\). 

The product-defining constraints are: 

\begin{subequations} \label{eq: product_duration_constraints}
\begin{align}
        q^{\text{p}} &\leq q^{\text{p}}_t  \quad & \label{eq: product_duration_constraints_1}\\ 
        q^{\text{p}}_t &= P_{t,\text{up}}^{\text{PCC}} - P_{t,\text{disp}}^{\text{PCC}}  \quad &\forall t \in \mathcal{T} \label{eq: product_duration_constraints_2}\\
        q^{\text{p}}_t &= P_{t,\text{disp}}^{\text{PCC}} - P_{t,\text{down}}^{\text{PCC}}.  \quad & \label{eq: product_duration_constraints_3}
\end{align}
\end{subequations}
The electrical quantities are mapped to the corresponding product quantity~\eqref{eq: product_duration_constraints_1}, and it is ensured that the product quantity is available for each time step during the delivery period~\eqref{eq: product_duration_constraints_2}-\eqref{eq: product_duration_constraints_3}. The product quantity \(q^{\text{p}}\) is defined as the difference between the power exchanged at the point of common coupling \(P_{t,s}^{\text{PCC}}\) in the dispatching state and the upward~\eqref{eq: product_duration_constraints_2} and/or downward~\eqref{eq: product_duration_constraints_3} activated reserve states.
If only upward or downward flexibility is considered, one of constraints~\eqref{eq: product_duration_constraints_2}-\eqref{eq: product_duration_constraints_3} must be omitted from the formulation. 

\textit{Network constraints.} The network is modeled using the linear DistFlow equations~\cite{19266}. This formulation leverages the radial topology assumed for the VPP distribution network to provide a direct linearized version of the power flow equations. Compared to alternative linearization approaches, such as the DC power flow model, the linear DistFlow approximation allows the estimation of voltage magnitudes and works well under large resistance-reactance ratios~\cite{19266}. 
In linear DistFlow, the power flow equations are expressed as:

\begin{subequations} \label{eq: lin_dist_flow_constraints}
\begin{align}
        &\sum_{(jk) \in \mathcal{E}_j^{\text{out}}} p_{jk,t,s} = \sum_{(ij) \in \mathcal{E}_j^{\text{in}}} p_{ij,t,s} + p_{j,t,s} \\
        &\sum_{(jk) \in \mathcal{E}_j^{\text{out}}} q_{jk,t,s} = \sum_{(ij) \in \mathcal{E}_j^{\text{in}}} q_{ij,t,s} + q_{j,t,s} \\
        &\hspace{4.2cm} \forall j  \in \mathcal{N}, \forall t \in \mathcal{T}, \forall s \in \mathcal{S} \nonumber\\
        &v_{j,t,s} - v_{k,t,s} = 2(r_{jk} p_{jk,t,s} + x_{jk} q_{jk,t,s}) \\
        &\hspace{3.9cm} \forall (jk)  \in \mathcal{E}, \forall t \in \mathcal{T}, \forall s \in \mathcal{S} \nonumber
\end{align}
\end{subequations}
where \(p_{j,t,s}\) and \(q_{j,t,s}\) are the active and reactive power injections in node \(j\) at time step \(t\), for state \(s\), respectively. Moreover, \(p_{jk,t,s}\) and \(q_{jk,t,s}\) are the real and reactive power flows in branch \(jk\), respectively. Finally, \(v_{j,t,s}\) is the squared voltage magitude at bus \(j\), \(r_{jk}\) and \(x_{jk}\) are the resistance and reactance of branch \(jk\), respectively. The sets include \( \mathcal{N} \) as the set of nodes, \( \mathcal{E} \) as the set of branches, \( \mathcal{E}_j^{\text{out}} \) as the set of branches exiting node \(j\), and \( \mathcal{E}_j^{\text{in}} \) as the set of branches entering node \(j\). Note that in a radial network \( |\mathcal{E}_j^{\text{in}}| = 1 \).

The voltage constraints are detailed in~\eqref{eq:voltage_limits_time_subs}. Branch flow constraints are formulated in~\eqref{eq:piecewise_linear_constraint} using a piecewise linearization to maintain a linear formulation:
\begin{subequations} \label{eq:voltage and current limits}
\begin{align}
        &(v_{j}^{\text{min}})^2 \leq v_{j,t,s} \leq (v_{j}^{\text{max}})^2 \label{eq:voltage_limits_time_subs} \\ 
        & \hspace{4.0cm} \forall j \in \mathcal{N}, \forall t \in \mathcal{T}, \forall s \in \mathcal{S} \nonumber\\
        & p_{jk,t,s} \cos\left(\frac{2\pi l}{N^{\text{seg}}}\right) + q_{jk,t,s} \sin\left(\frac{2\pi l}{N^{\text{seg}}}\right) \leq s_{jk}^{\max} \label{eq:piecewise_linear_constraint}\\
        & \hspace{4.2cm} \forall l \in \{0, \ldots, N^{\text{seg}} - 1\}, \nonumber \\
        & \hspace{3.7cm} \forall (jk) \in \mathcal{E}, \forall t \in \mathcal{T}, \forall s \in \mathcal{S}.  \nonumber
\end{align}
\end{subequations}
Here, \(v_{j}^{\text{min}}\) and \(v_{j}^{\text{max}}\) are the voltage limits of bus \(j\), \(s_{jk}^{\max}\) is the power capacity of branch \(jk\), and \(N^{\text{seg}}\) is the number of linearization segments.

The nodal active and reactive power injections are linked to the power exchanges of the DERs through the nodal balance constraints in~\eqref{eq:nodal_balance_P}~and~\eqref{eq:nodal_balance_Q}.
DERs can be categorized into four technology classes: DG, HP, BESS, and EV. Considering these technologies, the nodal power balance constraints are defined as:

\begin{multline}
        \sum_{i \in \mathcal{G}_j} P_{i,t,s}^{\text{G}} - \sum_{i \in \mathcal{HP}_j} P_{i,t,s}^{\text{HP}} + \sum_{i \in \mathcal{ST}_j} P_{i,t,s}^{\text{ST}} \\
        + \sum_{i \in \mathcal{EV}_j} P_{i,t,s}^{\text{EV}} - \sum_{i \in \mathcal{L}_j} P_{i,t}^{\text{L}} = p_{j,t,s} \cdot S^{\text{base}}\\
          \forall j \in \mathcal{N}, \forall t \in \mathcal{T}, \forall s \in \mathcal{S}
    \label{eq:nodal_balance_P} 
\end{multline}
\begin{multline}
        \sum_{i \in \mathcal{G}_j} Q_{i,t,s}^{\text{G}} - \sum_{i \in \mathcal{HP}_j} Q_{i,t,s}^{\text{HP}} + \sum_{i \in \mathcal{ST}_j} Q_{i,t,s}^{\text{ST}} \\
        + \sum_{i \in \mathcal{EV}_j} Q_{i,t,s}^{\text{EV}} - \sum_{i \in \mathcal{L}_j} Q_{i,t}^{\text{L}} = q_{j,t,s} \cdot S^{\text{base}} \\
        \quad \forall j \in \mathcal{N}, \forall t \in \mathcal{T}, \forall s \in \mathcal{S} 
    \label{eq:nodal_balance_Q}
\end{multline}
where \(P_{i,t,s}^{\text{G}}\), \(Q_{i,t,s}^{\text{G}}\), \(P_{i,t,s}^{\text{HP}}\), \(Q_{i,t,s}^{\text{HP}}\), \(P_{i,t,s}^{\text{ST}}\), \(Q_{i,t,s}^{\text{ST}}\), \(P_{i,t,s}^{\text{EV}}\), \(Q_{i,t,s}^{\text{EV}}\) are the active and reactive power exchanges of controllable DGs, HPs, BESS, and EVs. All non-controllable resources power withdrawals are aggregated in the \(P_{i,t}^{\text{L}}\) and \(Q_{i,t}^{\text{L}}\) parameters. Additionally, \(\mathcal{G}_j\), \(\mathcal{ST}_j\), \(\mathcal{EV}_j\), \(\mathcal{HP}_j\), and \(\mathcal{L}_j\) are the sets of resources connected to bus \(j\).

\textit{DERs' constraints}. The constraints modeling the operations of DERs are reported in~\eqref{eq: PV_flexibility_maximization_constraints}-\eqref{eq: EV_flexibility_maximization_constraints}. The DG constraints model the operations of renewable units:
\begin{subequations} \label{eq: PV_flexibility_maximization_constraints}
\begin{align}
        &0 \leq P_{i,t,s}^{\text{G}} \leq P_{i}^{\text{G,nom}} \cdot CF_{i,t}^{\text{G}} \label{eq: DG_P_limits}\\
        &-Q_{i}^{\text{G,nom}} \leq Q_{i,t,s}^{\text{G}} \leq Q_{i}^{\text{G,nom}} \label{eq: DG_Q_limits}\\
        & \hspace{4.2cm} \forall i \in \mathcal{G}, \forall t \in \mathcal{T}, \forall s \in \mathcal{S} \nonumber\\
        &- r^{\text{p}} \cdot RR_{i}^{\text{G,-}} \leq P_{i,t,\text{disp}}^{\text{G}} - P_{i,t,s}^{\text{G}} \leq r^{\text{p}} \cdot RR_{i}^{\text{G,+}} \label{eq:PV_ramp_constraint} \\
        & \hspace{3.0cm} \forall i \in \mathcal{G}, \forall t \in \mathcal{T}, \forall s \in \{\mathcal{S} - \text{disp}\} \nonumber 
\end{align}
\end{subequations}
where \(\mathcal{G}\) is the set of controllable generators in the VPP. A squared capability chart is assumed~\cite{Riaz2022}, and modeled with constraints~\eqref{eq: DG_P_limits}-\eqref{eq: DG_Q_limits}. The reactive power limit \( Q_{i}^{\text{G,nom}} \) is fixed based on inverter characteristics. The active power \( P_{i,t,s}^{\text{G}} \) can take any value between zero and a maximum, determined by the plant's nominal power \( P_{i}^{\text{G,nom}} \) and the capacity factor \( CF_{i,t}^{\text{G}} \), which models the availability of the primary energy source. Lastly, the ramp time constraint~\eqref{eq:PV_ramp_constraint} ensures that the difference between the dispatching and activated-reserve states does not exceed the generator's ramping capability (\(RR_{i}^{\text{G,-}}\), \(RR_{i}^{\text{G,+}}\)) during the product ramp time.

The HPs' operations are modeled as follows:
\begin{subequations} \label{eq: HP_flexibility_maximization_constraints}
\begingroup
\begin{align}
    & P_{i}^{\text{HP,min}} \cdot \delta_{i,t,s}^{\text{HP}} \leq P_{i,t,s}^{\text{HP}} \leq P_{i}^{\text{HP,max}} \cdot \delta_{i,t,s}^{\text{HP}} \label{eq: HP_P_limits}\\
    & Q_{i}^{\text{HP,min}} \leq Q_{i,t,s}^{\text{HP}} \leq Q_{i}^{\text{HP,max}} \label{eq: HP_Q_limits}\\
    & T_{i,t,s} = T_{i,t_{-1},s} + \frac{P_{i,t,s}^{\text{HP}} \cdot \Delta t \cdot \text{COP}_i}{C_{i}^{\text{th}}} \nonumber \\
    & \hspace{4cm}- \frac{T_{i,t_{-1},s} - T_{i, t}^{\text{amb}}}{R_{i}^{\text{th}} \cdot C_{i}^{\text{th}}} \cdot \Delta t \label{eq:HP_thermal_balance_constraint} \\
    & T_{i}^{\text{min}} \leq T_{i,t,s} \leq T_{i}^{\text{max}} \label{eq:HP_max_min_T_constraint}\\
    & \hspace{4cm} \forall i \in \mathcal{HP}, \forall t \in \mathcal{T}, \forall s \in \mathcal{S} \nonumber\\
    & T_{i,t_0,s} = T_{i,t_{|\mathcal{T}|},s} = T_{i}^{\text{target}} \label{eq:HP_cicle_contraint}\\
    & \hspace{5.2cm} \forall i \in \mathcal{HP}, \forall s \in \mathcal{S} \nonumber\\
    & - r^{\text{p}} \cdot RR_{i}^{\text{HP,-}} \leq P_{i,t,\text{disp}}^{\text{HP}} - P_{i,t,s}^{\text{HP}} \leq r^{\text{p}} \cdot RR_{i}^{\text{HP,+}} \label{eq:HP_ramp_constraint} \\
    & \hspace{2.7cm} \forall i \in \mathcal{HP}, \forall t \in \mathcal{T}, \forall s \in \{\mathcal{S} - \text{disp}\} \nonumber
\end{align}
\endgroup
\end{subequations}
where \(\mathcal{HP}\) denotes the set of HPs in the VPP, and \( \delta_{i,t,s}^{\text{HP}} \) is a binary variable representing their operational status. If \( \delta_{i,t,s}^{\text{HP}}=1 \), the HP is operating, else it is shut down. The real power \( P_{i,t,s}^{\text{HP}} \) and reactive power \( Q_{i,t,s}^{\text{HP}} \) outputs are constrained in~\eqref{eq: HP_P_limits}-\eqref{eq: HP_Q_limits} by \( P_{i}^{\text{HP,min}} \), \( P_{i}^{\text{HP,max}} \), \( Q_{i}^{\text{HP,min}} \), and \( Q_{i}^{\text{HP,max}} \), respectively. \( \text{COP}_i \) is the coefficient of performance. The indoor temperature \( T_{i,t,s} \) is determined based on the indoor temperature in the previous time step \(T_{i,t_{-1},s}\), the ambient temperature \( T_{i,t}^{\text{amb}} \), thermal capacitance \( C_{i}^{\text{th}} \), thermal resistance \( R_{i}^{\text{th}} \), and heat inflow of the considered building as depicted by~\eqref{eq:HP_thermal_balance_constraint}, modeling the dynamics of a first-order RC thermal circuit~\cite{amara2015comparison}. Constraint~\eqref{eq:HP_max_min_T_constraint} ensures that \( T_{i,t,s} \) remains within comfort bounds \( T_{i}^{\text{min}} \) and \( T_{i}^{\text{max}} \). Initial and final temperatures must meet the target temperature \( T_{i}^{\text{target}} \) to guarantee resource availability in the subsequent time steps~\eqref{eq:HP_cicle_contraint}. Lastly,~\eqref{eq:HP_ramp_constraint} models HP' ramping capabilities \( RR_{i}^{\text{HP,+}} \) and \( RR_{i}^{\text{HP,-}} \).

The EVs' operations are modeled as follows:
\begin{subequations} \label{eq: EV_flexibility_maximization_constraints}
\begingroup
\allowdisplaybreaks
\begin{align}
    &0 \leq P_{i,t,s}^{\text{EV,ch}} \leq P_{i}^{\text{EV,max,ch}} \cdot \delta_{i,t,s}^{\text{EV}} \cdot K^{\text{EV}}_{i,t} \label{eq: EV_P_charge_constraint}\\
    &0 \leq P_{i,t,s}^{\text{EV,dis}} \leq P_{i}^{\text{EV,max,dis}} \cdot (1 - \delta_{i,t,s}^{\text{EV}}) \cdot K^{\text{EV}}_{i,t} \\
    &P_{i,t,s}^{\text{EV}} = P_{i,t,s}^{\text{EV,dis}} - P_{i,t,s}^{\text{EV,ch}}\\
    &Q_{i}^{\text{EV,min}} \cdot K^{\text{EV}}_{i,t} \leq Q_{i,t,s}^{\text{EV}} \leq Q_{i}^{\text{EV,max}} \cdot K^{\text{EV}}_{i,t} \label{eq: EV_Q_constraint}\\
    &\text{SOC}_{i,t,s}^{\text{EV}} = \text{SOC}_{i,t_{-1},s}^{\text{EV}}
      - \frac{P_{i,t,s}^{\text{EV,dis}} \cdot \Delta t}{\eta_{i}^{\text{EV,dis}} \cdot C_{i}^{\text{EV}}} \nonumber\\
    &\hspace{4cm} + \frac{P_{i,t,s}^{\text{EV,ch}} \cdot \Delta t \cdot \eta_{i}^{\text{EV,ch}}}{C_{i}^{\text{EV}}} \label{eq:EV_energy_balance_constraint}\\
    &\text{SOC}_{i}^{\text{EV,min}} \leq \text{SOC}_{i,t,s}^{\text{EV}} \leq \text{SOC}_{i}^{\text{EV,max}} \label{eq:EV_max_min_SOC_constraint}\\
    &\hspace{4cm} \forall i \in \mathcal{EV}, \forall t \in \mathcal{T}, \forall s \in \mathcal{S} \nonumber\\
    &\text{SOC}_{i,t_0,s}^{\text{EV}} = \text{SOC}_{i}^{\text{EV,ini}} \label{eq:EV_initial_SOC_constraint}\\
    &\text{SOC}_{i,t_{|\mathcal{T}|},s}^{\text{EV}} \geq \text{SOC}_{i}^{\text{EV,ini}} + R_{i}^{\text{EV,ch,min}} \cdot \Delta t \cdot \sum_{t} K^{\text{EV}}_{i,t} \label{eq:EV_final_SOC_constraint}\\
    &\hspace{5.2cm}\forall i \in \mathcal{EV}, \forall s \in \mathcal{S} \nonumber\\
    &- r^{\text{p}} \cdot RR_{i}^{\text{EV},-} \leq P_{i,t,\text{disp}}^{\text{EV}} - P_{i,t,s}^{\text{EV}} \leq r^{\text{p}} \cdot RR_{i}^{\text{EV},+} \label{eq:EV_ramp_constraint}\\
    &\hspace{2.7cm}\forall i \in  \mathcal{EV}, \forall t \in \mathcal{T}, \forall s \in \{\mathcal{S} - \text{disp}\} \nonumber
\end{align}
\endgroup
\end{subequations}
where EVs are modeled through their charging events, and \(\mathcal{EV}\) is the set of EVs charging events in the considered time horizon. Each event is characterized by start date, arrival state of charge \(\text{SOC}_{i}^{\text{EV,ini}}\), duration, maximum charging power \(P_{i}^{\text{EV,max,ch}}\), maximum discharging power \(P_{i}^{\text{EV,max,dis}}\), required minimum charge rate \(R_{i}^{\text{EV,ch,min}}\), and reactive power limits \(Q_{i}^{\text{EV,min}}\) and \(Q_{i}^{\text{EV,max}}\). Constraints~\eqref{eq: EV_P_charge_constraint}-\eqref{eq: EV_Q_constraint} enforce active-reactive power limits, and define the net active power exchange \(P_{i,t,s}^{\text{EV}}\). Power exchange is allowed in both directions, accounting for vehicle-to-grid operation~\cite{6099519}. The binary variable \(\delta_{i,t,s}^{\text{EV}}\) is added to prevent simultaneous charging and discharging. Additionally, the parameter \( K^{\text{EV}}_{i,t} \) is introduced, which equals 1 when the EV is connected to the grid and zero otherwise. It is derived by intersecting the charging event's time interval with the product delivery period \( \mathcal{T} \). The energy balance constraint~\eqref{eq:EV_energy_balance_constraint} connects the SOC at each time step \(\text{SOC}_{i,t,s}^{\text{EV}}\) to the SOC in the previous time step \(\text{SOC}_{i,t_{-1},s}^{\text{EV}}\), considering the charging and discharging power (\(P_{i,t,s}^{\text{EV,ch}}\), \(P_{i,t,s}^{\text{EV,dis}}\)), efficiencies (\(\eta_{i}^{\text{EV,ch}}\), \(\eta_{i}^{\text{EV,dis}}\)), and the battery capacity \(C_{i}^{\text{EV}}\). Moreover, the SOC is constrained in~\eqref{eq:EV_max_min_SOC_constraint} within the interval [\(\text{SOC}_{i}^{\text{EV,min}}\), \(\text{SOC}_{i}^{\text{EV,max}}\)] to guarantee a minimum charge at any time. To further ensure a minimum service level, the initial SOC is set to the vehicle's arrival value~\eqref{eq:EV_initial_SOC_constraint}, whereas the final SOC is set to the arrival value plus the required minimum charge rate~\eqref{eq:EV_final_SOC_constraint}. Finally,~\eqref{eq:EV_ramp_constraint} describes the EV ramp time constraints.

The constraints of BESS are formulated analogously to the EV constraints in~\eqref{eq: EV_flexibility_maximization_constraints}. The only differences are that 1) storage systems are always connected, and 2) the SOC in the beginning must be equal to the SOC at the end of the observation period. The BESS constraints are therefore omitted here.

\textit{Uncertain parameters}. The generators' capacity factors, the ambient temperatures, the EV presence parameters, and the non-dispatchable load powers are subject to forecasting uncertainty because the power reserve is scheduled \(t^{\text{lead}}\) ahead of delivery. Thus, they define the uncertainty set \(\mathcal{P}^{\text{u},\text{mfa}} = \{ CF_{i,t}^{\text{G}}, T_{i, t}^{\text{amb}}, K^{\text{EV}}_{i,t}, P_{i,t}^{\text{L}}, Q_{i,t}^{\text{L}}\}\).

\subsubsection{Subset simulation}
\label{sec: method subset simulation}
In this section, we first formulate the maximum flexibility assessment model as a reliability problem. Then, we present SS methodology and its adaptation to extreme quantile estimation. Finally, we assess the accuracy of the SS quantile estimator.

\textit{Problem reformulation.}
The flexibility maximization problem is treated as the performance function \(g(\boldsymbol{z})\), which, given the uncertain parameters \(\boldsymbol{z}=(z_1,\ldots,z_u) \in \mathcal{P}^{\text{u},\text{mfa}}\) outputs the system response, which here coincides with the maximum product quantity.
The system is in a failure state when the maximum product quantity \(g(\boldsymbol{z})\) that the VPP can provide is lower than \(q^{\text{p,max}}\). The failure domain \(F\) is the fraction of the uncertain input space for which the system is in a failure state.
With \(\pi(\boldsymbol{z})\) denoting the joint probability density function (PDF) of \(\boldsymbol{z}\), the failure probability \(p^F\) can be defined as:
\begin{equation}
    p^F = \mathbb{P}\left( \boldsymbol{z} \in F \right) = \int_F \pi \left( \boldsymbol{z} \right) d\boldsymbol{z}.
    \label{eq:failure_probability}
\end{equation}
The \(\alpha^{\text{p}}\)-quantile estimation problem translates into finding the value of \(q^{\text{p,max}}\), which determines a failure domain \(F\), such that \(p^F = \alpha^{\text{p}}\). 

\textit{Subset simulation for quantile estimation.} In its traditional formulation, SS breaks down the rare-event probability estimation problem into a series of simpler subproblems in which the probability of more likely events is estimated. In this context, we modify SS to estimate extreme quantiles. Instead of directly computing the extreme quantile associated with a small probability \(\alpha^{\text{p}}\), we reformulate the problem into a sequence of \(N^{\text{L}}\) nested quantile estimation subproblems, called SS levels, in which intermediate quantiles \(q^{\text{p,*}}_{k}\) with a larger probability \(\alpha^\text{ss} = (\alpha^{\text{p}}) ^{1/{N^{\text{L}}}}\) are estimated. Here, \(k\) is the index of the subset level, with \(k \in [1, N^{\text{L}}]\). 

The procedure begins at level 1 by applying DMC to determine the first intermediate quantile \(q^{\text{p,*}}_1\) corresponding to failure domain \(F_1\) such that its probability equals the new target value \(\alpha^\text{ss}\). Then, given \(F_1\), the conditional distribution \(\pi(\boldsymbol{z} \mid \boldsymbol{z} \in F_1)\) is sampled using Markov Chain Monte Carlo (MCMC) and the modified Metropolis algorithm~\cite{doi:https://doi.org/10.1002/9781118398050.ch5}. This allows the estimation of the next intermediate quantile \(q^{\text{p,*}}_2\), defining a new failure domain \(F_2\), nested within \(F_1\), with probability \(\alpha^\text{ss}\) on the conditional distribution and a probability \((\alpha^\text{ss})^2\) on the original distribution. For each subsequent level \(k\), the intermediate quantile \(q^{\text{p,*}}_{k}\) is determined such that \(F_{k} = \{\boldsymbol{z} \mid g(\boldsymbol{z}) < q^{\text{p,*}}_{k}\}\) within \(F_{k-1}\) has a probability \(\mathbb{P}(F_{k} \mid F_{k-1}) = \alpha^\text{ss}\). The SS procedure ends when \(k = N^{\text{L}}\). Here, the probability of \(F_{N^{\text{L}}}\) on the original distribution equals the target quantile probability \(\alpha^{\text{p}}\):
\begin{equation} \label{eq:probability product}
\begin{aligned}
     \mathbb{P}(F_{N^{\text{L}}}) &= \mathbb{P}(F_1) \cdot \mathbb{P}(F_2 \mid F_1) \cdot \ldots \cdot \mathbb{P}(F_{N^{\text{L}}} \mid F_{{N^{\text{L}}}-1}) \\
     &= (\alpha^\text{ss})^{N^{\text{L}}} = \alpha^{\text{p}}    
\end{aligned}
\end{equation}
therefore, the \({N^{\text{L}}}\)-th intermediate quantile \(q^{\text{p,*}}_{N^{\text{L}}}\) corresponds to the extreme quantile of interest \(q^{\text{p,max}} = q^{\text{p,*}}_{N^{\text{L}}}\).

\textit{Accuracy assessment.} To assess the result's accuracy, a formula to compute the single-run COV of the SS quantile estimator is proposed. According to~\cite{doi:https://doi.org/10.1002/9781118398050.ch5}, the COV of the quantile estimator is linked to the COV of the probability estimator as:
\begin{equation}
    \text{COV}^{\text{quant}} = \frac{\text{COV}^{\text{prob}} \cdot \alpha^{\text{p}}}{f_{Q^{\text{p}}}(q^{\text{p,max}}) \cdot q^{\text{p,max}}}
    \label{eq: COV_quantile_expression}
\end{equation}
where \(f_{Q^{\text{p}}}(q^{\text{p,max}})\) represents the model's response PDF evaluated at the target quantile, and \(\text{COV}^{\text{prob}}\) is the COV of the corresponding naive SS probability estimator. While \(\alpha^{\text{p}}\) is known and \(q^{\text{p,max}}\) is the result of the SS procedure, \(\text{COV}^{\text{prob}}\) and \(f_{Q^{\text{p}}}(q^{\text{p,max}})\) have to be determined. In particular, \(\text{COV}^{\text{prob}}\) can be computed from the SS samples with the procedure presented in~\cite{doi:https://doi.org/10.1002/9781118398050.ch5}, and \(f_{Q^{\text{p}}}(q^{\text{p,max}})\) is computed from the available samples as described in~\cite{dong_quantile_2017}.

\subsection{Flexibility cost assessment model}
\label{sec: flexibility cost assessment model}

The flexibility cost assessment model quantifies the supply curve of the power reserve capacity product.
The product cost is characterized as a function of the product quantity using the \(\epsilon\)-constrained approach~\cite{4308298}. The quantity range from zero to the maximum \(q^{\text{p,max}}\) is divided into a set of \(N^{\text{step}}\) equally-spaced steps, resulting in \(N^{\epsilon} = N^{\text{step}} + 1\) target product quantity values \(q^{\text{p,}\epsilon}_n\), with \(n \in \{0, \ldots, N^{\text{step}}\}\). For each value \(q^{\text{p,}\epsilon}_n\), the risk-adjusted product cost \(c^{\text{p},\alpha_{\text{c}}}(q^{\text{p,}\epsilon}_n)\) is determined, modeling the impact of forecast uncertainty on product cost and the VPP manager's risk attitude.

Specifically, for each \( q^{\text{p,}\epsilon}_n \) and for each realization of the uncertain parameters \(\boldsymbol{z} \in \mathcal{P}^{\text{u},\text{fca}}\), we compute the product cost \(c^{\text{p}}(q^{\text{p,}\epsilon}_n,\boldsymbol{z})\) by solving an operational optimization problem, which finds the network’s and DERs’ operational state that minimizes total VPP operations cost \(C(\boldsymbol{x},\boldsymbol{y},\boldsymbol{z},q^{\text{p,}\epsilon}_n)\). This is formulated as a MILP: 
\begin{equation}
    \begin{aligned}
    & \underset{\boldsymbol{x},\boldsymbol{y}}{\text{min}} & & C(\boldsymbol{x},\boldsymbol{y},\boldsymbol{z},q^{\text{p,}\epsilon}_n) \\
    & \text{s.t.} & & \text{Product-defining, network, DER's constraints} \\
    & & & q^{\text{p}}(\boldsymbol{x},\boldsymbol{y},\boldsymbol{z}) \geq q^{\text{p,}\epsilon}_n \\
    & & & \text{Cost-defining constraints.}
    \end{aligned}
\label{eq: method cost minimization problem}
\end{equation}
In addition to the constraints introduced in the flexibility maximization model~\eqref{eq: method flexibility maximization model}, the cost minimization model includes the minimum product quantity constraint \(q^{\text{p}}(\boldsymbol{x},\boldsymbol{y},\boldsymbol{z}) \geq q^{\text{p,}\epsilon}_n\), introduced to apply the \(\epsilon\)-constrained method, and the cost-defining constraints. 
The products's cost is computed as the dual variable of the minimum product quantity constraint~\cite{7530870}:
\begin{equation}
    c^{\text{p}}(q^{\text{p,}\epsilon}_n,\boldsymbol{z}) = \lambda^{\text{q}}: q^{\text{p}}(\boldsymbol{x},\boldsymbol{y},\boldsymbol{z}) \geq q^{\text{p,}\epsilon}_n.
\end{equation}
Since the cost is subject to uncertainty in \(\boldsymbol{z}\), the VPP manager accounts for risk by selecting a cost estimate based on a quantile of the cost distribution:
\begin{equation}
    c^{\text{p},\alpha_{\text{c}}}(q^{\text{p,}\epsilon}_n) = \inf \left\{ \tilde{c^{\text{p}}} \in \mathbb{R} \mid \mathbb{P}(c^{\text{p}}(q^{\text{p,}\epsilon}_n, \boldsymbol{z}) \leq \tilde{c^{\text{p}}}) = \alpha^{\text{c}} \right\}.
\end{equation}
The quantile probability \( \alpha^{\text{c}} \) reflects the manager's risk attitude, influencing the price at which the product is offered. Higher values of \( \alpha^{\text{c}} \) correspond to more conservative pricing decisions. The quantile is found through DMC with Latin hypercube sampling (LHS), as \( \alpha^{\text{c}} \) is not usually extreme.
By solving for all target quantities \(q^{\text{p}\epsilon}_n\), the flexibility cost assessment model constructs the risk-adjusted supply curve \(c^{\text{p},\alpha_{\text{c}}}(q^{\text{p,}\epsilon}_n)\).

The remainder of this section details the cost-defining constraints introduced in \eqref{eq: method cost minimization problem} and defines the uncertain parameters' set \(\mathcal{P}^{\text{u},\text{fca}}\). 

\textit{Cost-defining constraints}. The operational costs \(C\) are computed as the sum of the DER costs and energy costs:
\begin{equation}
C =  C^{\text{G}} + C^{\text{HP}} + C^{\text{EV}} + C^{\text{ST}} + C^{\text{E}}.
\label{eq: operations_cost_definition}
\end{equation}
The generation cost \(C^{\text{G}}\) is defined considering fixed marginal costs for the active and reactive power as follows:
\begin{subequations} \label{eq:Q_splitting_constraint_generators} 
\begin{equation}
    C^{\text{G}} = \sum_{t \in \mathcal{T}} \sum_{i \in \mathcal{G}} \left( P^{\text{G}}_{i,t,\text{disp}} \cdot c^{\text{G,P}}_i + \left( Q^{\text{G},+}_{i,t,\text{disp}}  + Q^{\text{G},-}_{i,t,\text{disp}} \right) \cdot c^{\text{G,Q}}_i \right)
    \label{eq:generator_cost_constraint} \\
\end{equation}
\begin{align}
        Q^{\text{G}}_{i,t,\text{disp}} &= Q^{\text{G},+}_{i,t,\text{disp}} - Q^{\text{G},-}_{i,t,\text{disp}} & \label{eq:Q_splitting_constraint_generators_1}\\
        Q^{\text{G},+}_{i,t,\text{disp}} &, Q^{\text{G},-}_{i,t,\text{disp}} \geq 0 & \quad \forall i \in \mathcal{G}, \forall t \in \mathcal{T}
\end{align}
\end{subequations}
where \(Q^{\text{G},-}_{i,t,\text{disp}}\) and \(Q^{\text{G},+}_{i,t,\text{disp}}\) are the reactive power withdrawals and injections, respectively. Moreover, \(c^{\text{G,P}}_i\) and \(c^{\text{G,Q}}_i\) are the real and reactive power marginal costs, respectively. 

The heat pumps' cost \(C^{\text{HP}}\) is defined as:
\begin{subequations} \label{eq:Q_minus_positivity_heat_pumps}
\begin{multline}
    C^{\text{HP}} = \sum_{t \in \mathcal{T}} \sum_{i \in \mathcal{HP}} \left( \left( Q^{\text{HP},+}_{i,t,\text{disp}} + Q^{\text{HP},-}_{i,t,\text{disp}} \right) \cdot c^{\text{HP,Q}}_{i} \right.\\
    \left. + \left( \Delta T^{+}_{i,t} + \Delta T^{-}_{i,t} \right) \cdot c^{\text{T}}_{i} \right)
\label{eq: heat_pumps_cost_constraint}
\end{multline}
\begin{align}
& Q^{\text{HP}}_{i,t,\text{disp}} = Q^{\text{HP},+}_{i,t,\text{disp}} - Q^{\text{HP},-}_{i,t,\text{disp}} & \label{eq:eq:Q_minus_positivity_heat_pumps_1}\\
& Q^{\text{HP},+}_{i,t,\text{disp}}, Q^{\text{HP},-}_{i,t,\text{disp}} \geq 0 & \\
& T_{i,t,\text{disp}} - T^{\text{target}}_{i} = \Delta T^{+}_{i,t} - \Delta T^{-}_{i,t} &\\
& \Delta T^{+}_{i,t}, \Delta T^{-}_{i,t} \geq 0 &  \forall i \in \mathcal{HP}, \forall t \in \mathcal{T} \label{eq:eq:Q_minus_positivity_heat_pumps_6}
\end{align}
\end{subequations}
where \(Q^{\text{HP},-}_{i,t,\text{disp}}\) and \(Q^{\text{HP},+}_{i,t,\text{disp}}\) are the reactive power withdrawals and injections, respectively, and \(\Delta T^{+}_{i,t}\) and \(\Delta T^{+}_{i,t}\) are the positive and negative temperature deviations from \(T^{\text{target}}_{i}\), respectively. Fixed reactive power marginal cost \(c^{\text{HP,Q}}_i\) and a temperature discomfort penalty \(c^{\text{T}}_{i}\) are included.

The cost of electric vehicles' operations \(C^{\text{EV}}\) is defined as:
\begin{subequations} \label{eq:Q_plus_positivity_electric_vehicles} 
\begin{multline}
    C^{\text{EV}} = \sum_{t \in \mathcal{T}} \sum_{i \in \mathcal{EV}} \left( \left( P^{\text{EV},\text{ch}}_{i,t,\text{disp}} + P^{\text{EV},\text{dis}}_{i,t,\text{disp}} \right) \cdot c^{\text{EV,P}}_{i} \right. \\
    \left. + \left( Q^{\text{EV},+}_{i,t,\text{disp}} + Q^{\text{EV},-}_{i,t,\text{disp}} \right) \cdot c^{\text{EV,Q}}_{i}  \right)
\label{eq:electric_vehicles_cost_constraint}
\end{multline}
\begin{align}
Q_{i,t,\text{disp}}^{\text{EV}} &= Q_{i,t,\text{disp}}^{\text{EV},+} - Q_{i,t,\text{disp}}^{\text{EV},-} & \label{eq:Q_plus_positivity_electric_vehicles_1}\\
Q_{i,t,\text{disp}}^{\text{EV},+} &, Q_{i,t,\text{disp}}^{\text{EV},-}\geq 0& \quad \forall i \in \mathcal{EV}, \forall t \in \mathcal{T}  \label{eq:Q_plus_positivity_electric_vehicles_3}
\end{align}
\end{subequations}
where \(Q^{\text{EV},-}_{i,t,\text{disp}}\) and \(Q^{\text{EV},+}_{i,t,\text{disp}}\) are the reactive power withdrawals and injections, respectively, and \(c^{\text{EV,P}}_i\) and \(c^{\text{EV,Q}}_i\) are the fixed active and reactive power marginal costs, respectively. The real power marginal cost \(c^{\text{EV,P}}_{i}\) is used to model the cost of battery degradation~\cite{GORMAN2022105832, He2018}.

The costs of storage \(C^{\text{ST}}\) are defined using the same formulation~\eqref{eq:Q_plus_positivity_electric_vehicles} as for EVs.

The electricity cost \(C^{\text{E}}\) is defined as:
\begin{subequations}    \label{eq:metering_areas_withdrawal_positivity_constraint}
\begin{equation}
C^{\text{E}} = \sum_{t \in \mathcal{T}} \sum_{m \in \mathcal{MA}} \left( \left( P^{\text{inj}}_{m,t} - P^{\text{wit}}_{m,t} \right) \cdot c^{\text{market}}_t + P^{\text{wit}}_{m,t} \cdot c^{\text{tariff}}_t \right)
\label{eq: energy_cost_constraint}
\end{equation}
\begin{align}
&\sum_{i \in \mathcal{G}_m} P_{i,t,\text{disp}}^{\text{G}} + \sum_{i \in \mathcal{ST}_m} P_{i,t,\text{disp}}^{\text{ST}} + \sum_{i \in \mathcal{EV}_m} P_{i,t,\text{disp}}^{\text{EV}}  \nonumber\\
& \qquad - \sum_{i \in \mathcal{HP}_m} P_{i,t,\text{disp}}^{\text{HP}} - \sum_{i \in \mathcal{L}_m} P_{i,\text{disp}}^{\text{L}} = P^{\text{inj}}_{m,t} - P^{\text{wit}}_{m,t} 
\label{eq:metering_areas_withdrawal_positivity_constraint_1} \\ 
&P^{\text{inj}}_{m,t}, P^{\text{wit}}_{m,t} \geq 0 \qquad \qquad \qquad \forall m \in \mathcal{MA}, \forall t \in \mathcal{T} \label{eq:metering_areas_withdrawal_positivity_constraint_3}
\end{align}
\end{subequations}
where \(c^{\text{market}}_t\) is the day-ahead market (DAM) electricity price, and \(c^{\text{tariff}}_t\) denotes the network and system tariffs. Resources are organized into metering areas, each corresponding to a region served by a single meter. For each metering area \(m \in \mathcal{MA}\), the net power exchange is computed and split into power injections \(P^{\text{inj}}_{m,t}\) and withdrawals \(P^{\text{wit}}_{m,t}\) in~\eqref{eq:metering_areas_withdrawal_positivity_constraint_1}-\eqref{eq:metering_areas_withdrawal_positivity_constraint_3}. \(\mathcal{G}_m\), \(\mathcal{ST}_m\), \(\mathcal{EV}_m\), \(\mathcal{HP}_m\) and \(\mathcal{L}_m\) are the set of DERs within the \(m\)-th metering area. The DAM electricity price \(c^{\text{market}}_t\) is applied to both injections and withdrawals. Network and system tariffs \(c^{\text{tariff}}_t\) are only applied to withdrawals \(P^{\text{wit}}_{m,t}\), following typical European billing structures \cite{ACER2023}.

\textit{Uncertain parameters}. As the power reserve markets are typically cleared before the DAM, electricity prices \(c^{\text{market}}_t\) are unknown at \(t^{\text{lead}}\). Hence, the uncertain parameters set of the cost minimization problem is \(\mathcal{P}^{\text{u},\text{fca}} = \mathcal{P}^{\text{u},\text{mfa}} \cup \{ c^{\text{market}}_t \}\).

\section{Case study}
\label{sec:cstudy}
The following provides an overview of the considered case study, the uncertainty modeling, and the computational aspects of the implementation.

\subsection{Case study description}
\label{sec:cstudy_description}
\begin{figure}
    \centering
    \includegraphics[width=\linewidth]{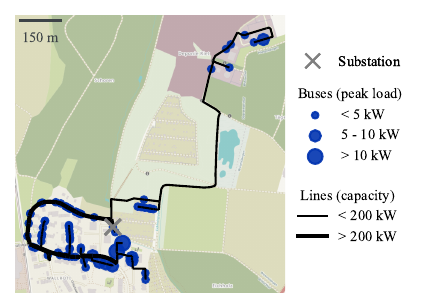}
    \caption{Case study network map showing the 97 buses, lines, and substation location.}
    \label{fig:network_map}
\end{figure}

The proposed framework is tested on a VPP based on a low-voltage network in northern Switzerland. This network is synthetically generated by~\cite{Oneto_2023} using open source data and shown in Fig.~\ref{fig:network_map}. It comprises 97 buses, with \(\SI{150}{kW}\) of peak non-dispatchable load.
The data form~\cite{Oneto_2023} includes the grid topology, line limits, and nodal non-dispatchable peak loads.
The penetration rates for DERs, HPs, and EVs are chosen to reflect the projected 2030 levels~\cite{EnergyPerspectives2050+}.
The network has installed about \(\SI{150}{kWp}\) of rooftop PV generation, \(\SI{85}{kW}\) of installed HPs, \(\SI{75}{kWh}\) of BESS, and 40 EV charging events during the considered time horizon. The EVs have an average battery capacity of \(\SI{70}{kWh}\) and an average maximum charging power of \(\SI{7}{kW}\).

The developed model is generic and can be applied to assess the provision of any power reserve capacity product. In this study, we consider a power reserve capacity product traded one day ahead of delivery, characterized by a duration of \(t^{\text{p}} = \SI{4}{h}\), from \(t^{\text{d}} = \SI{8}{am}\) to \(\SI{12}{pm}\). Furthermore, the product is symmetrical, with a ramp time requirement of \(r^{\text{p}} = \SI{5}{min}\) and a reliability requirement of \(R^{\text{p}} = \SI{99.9}{\%}\). The simulations assume a working day in May, with a time resolution \(\Delta t  = \SI{1}{h}\), leading to four time steps \(\mathcal{T} = \{8, 9, 10, 11 \}\). This product is similar to the secondary reserve products traded in most European markets, including the Swiss one~\cite{swissgrid2022_ASproducts, swissgrid2022_prequalification}.

Electricity prices \(c_t^{\text{market}}\) are based on the DAM prices of the Swiss market zone for the selected day in 2024~\cite{entsoe_transparency}.
Network and system tariffs \(c_t^{\text{tariff}}\) are set at \(\SI{206.5}{CHF/MWh}\)~\cite{ewz_2024_tariff}. Metering areas coincide with network nodes.

\subsection{Uncertainty modelling}
\label{sec:cstudy_uncertainty_modelling}

The uncertainty considered in this work relates to day-ahead forecasting errors of key parameters. The uncertain parameter set is outlined in Sections~\ref{sec: method flexibility maximization model} and~\ref{sec: flexibility cost assessment model} as \(\mathcal{P}^{\text{u}} = \mathcal{P}^{\text{u},\text{mfa}} \cup \mathcal{P}^{\text{u},\text{fca}} = \{ CF_{i,t}^{\text{G}}, T_{i, t}^{\text{amb}}, K^{\text{EV}}_{i,t}, P_{i,t}^{\text{L}}, Q_{i,t}^{\text{L}}, c^{\text{market}}_t \}\). The uncertainty model parameters are summarized in Table~\ref{tab:uncertainty_model}.

Due to the case study's limited spatial and temporal scope, each error is assumed to affect all resources equally at every time step. This allows the uncertainty to be modeled by five lumped error parameters: \(E^{\text{G}}\), \(E^{\text{T}}\), \(E^{\text{EV}}\), \(E^{\text{L}}\), and \(E^{\text{market}}\). A fixed power factor is adopted for non-dispatchable load. Thus, the uncertainty around \(P_{i,t}^{\text{L}}\) and \( Q_{i,t}^{\text{L}}\) can be described by one error parameter \(E^{\text{L}}\).

For electricity demand, solar irradiance, ambient temperature, and electricity price, forecasting errors are assumed to follow a normal distribution with zero mean~\cite{6520086, VANDERMEER20181484} and thus reflect unbiased estimators. 
The standard deviations are derived from a literature review of state-of-the-art forecasting tools. Errors are applied to reference profiles.

A different approach is used for EV schedules. The schedule from the mobility database~\cite{10694497} provides the upper bound (reference) for movements in the area. A schedule disruption parameter is introduced to account for the uncertainty in EV movements. A value of zero implies that the realized schedule matches the database. Values between zero and one indicate that a randomly selected proportion of scheduled events is removed equal to the disruption parameter. The disruption parameter is assumed to follow a uniform distribution from \(\SI{0}{}\) to \(\SI{20}{\%}\). 

\begin{table}
\centering
\caption{Uncertainty model parameters for the considered forecasting errors}
    \begin{tabular}{|c|c|c|c|c|}
    \hline
    \textbf{Error}        & \textbf{PDF}   & \boldmath\(\sigma\)              & \boldmath\(\mu\)        & \textbf{Ref.}                         \\ \hline
    \(E^{\text{L}}\)                & Normal         & \(\SI{10.8}{\%}\)               & 0                     &~\cite{8039509}                        \\ \hline
    \(E^{\text{G}}\)           & Normal         & \(\SI{8.15}{\%}\)                & 0                     &~\cite{solcast_forecast_accuracy}      \\ \hline
    \(E^{\text{T}}\)          & Normal         & \(\SI{1.50}{K}\)                 & 0                     &~\cite{meteoblue_weather_accuracy}     \\ \hline
    \(E^{\text{market}}\)     & Normal         & \(\SI{4.28}{\frac{CHF}{MWh}}\) & 0                     &~\cite{LAGO2021116983}                 \\ \hline
    \(E^{\text{EV}}\)           & Uniform        & \(\SI{5.77}{\%}\)                & \(\SI{10}{\%}\)         &    N.A.                                   \\ \hline
    \end{tabular}
\label{tab:uncertainty_model}
\end{table}
All five errors are assumed to be independent, although the authors acknowledge that some correlations exist, for instance, between temperature, irradiance, and electricity demand. However, this simplification is deemed acceptable for this case study, which aims to showcase the method.

The subset simulation in the maximum flexibility assessment model is set up with \(N^{\text{L}} = \SI{3}{}\) levels, a step probability of \(\alpha^\text{ss} = \SI{10}{\%}\), and 1000 samples per level. These parameters are chosen to achieve a coefficient of variation on the quantile estimator of \(\text{COV}^{\text{quant}} = \SI{1}{\%}\) according to~\eqref{eq: COV_quantile_expression}. 
In the flexibility cost assessment model, we chose \( N^{\text{step}} = \SI{30}{} \) \(\epsilon\)-constrained steps to ensure a dense characterization of the supply curve. For each step, the uncertainty analysis is conducted with \(N^{\text{dmc}} = \SI{1000}{}\) LHS samples to guarantee a COV below \(\SI{1}{\%}\) in the 5th and 95th percentile estimators~\cite{dong_quantile_2017}.

\subsection{Computational aspects}
\label{sec:cstudy_computational_aspects}
The flexibility assessment model is implemented in Python, using Pyomo~\cite{hart2011pyomo} as a parser and Gurobi~\cite{gurobi} as a solver. The model is executed on the ETH Zurich \textit{EULER} high-performance computing cluster. The sampling procedure is parallelized using MPI~\cite{10.5555/898758} on a node with 100 cores. The flexibility maximization problem with 11730 continuous variables, 620 binary variables, and 29100 constraints takes \(\sim \SI{5}{s}\) to solve. The cost minimization problem is slightly larger due to the addition of the cost-defining constraints and takes \(\sim \SI{7}{s}\) to solve.

\section{Results and discussion}
\label{sec:results}
This section presents the case study results. First, Section~\ref{sec: supply_curve} presents the supply curve of the selected reserve capacity product and discusses its shape and the computational advancements that are achieved when using the SS method.
Thereafter, Section~\ref{sec: impact_of_products_requirements} discusses the impact of the product's technical requirements on the VPP reserve provision capability.

\subsection{Reserve product supply curve}
\label{sec: supply_curve}
\begin{figure}
    \centering
    \includegraphics[width=\linewidth]{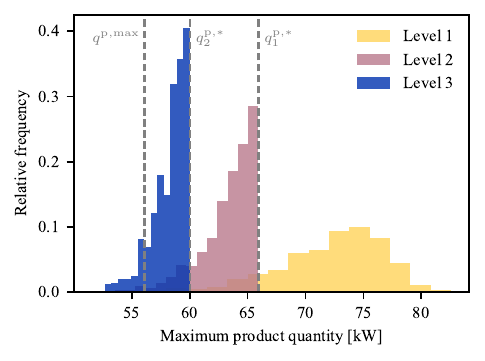}
    \caption{Maximum product quantity histograms for the three subset simulation levels. The intermediate quantiles are reported as \( q^{\text{p},\text{*}}_1 \) and \( q^{\text{p},\text{*}}_2 \), while \( q^{\text{p},\text{max}} \) is the target quantile that fulfils the reliability requirement.}
    \label{fig:subset_simulation_histogram}
\end{figure}

Fig.~\ref{fig:subset_simulation_histogram} shows the results of the maximum flexibility assessment for three SS levels, each consisting of 1000 samples, in the form of histograms and their intermediate quantiles. Level 1 corresponds to the initial DMC sampling, which determines the first intermediate quantile \( q^{\text{p},\text{*}}_1 \). Levels 2 and 3 represent the subsequent MCMC stages, where the algorithm samples the conditional distributions to estimate the intermediate quantile \( q^{\text{p},\text{*}}_2 \), and, ultimately, the target quantile \( q^{\text{p},\text{max}} \).
Thus, the maximum product quantity the VPP can reliably provide is estimated to be \(q^{\text{p,max}} = \SI{56.0}{kW}\). 

A total of \(N^{\text{ss}} = \SI{2800}{}\) samples were required to achieve the target COV for the quantile estimator. Only 2800 samples are needed instead of 3000 because, at each level, 100 samples are re-used as seeds for the MCMC chains of the subsequent level. The performance of the SS quantile estimator is compared with that of the DMC using LHS. The results are summarized in Table~\ref{table:quantile_estimation_comparison}. We find that achieving the same COV with the DMC quantile estimator requires 9000 samples, according to the COV estimation procedure presented in~\cite{dong_quantile_2017}. Therefore, the SS technique reduced the number of required samples (i.e., function evaluations) by \SI{69}{\%}, demonstrating its effectiveness for extreme quantile estimation.

\begin{table}[b]
\caption{Performance comparison between subset simulation (SS) and direct Monte Carlo (DMC). The coefficient of variation (COV) is estimated following \cite{dong_quantile_2017} for DMC and \eqref{eq: COV_quantile_expression} for SS.}
\centering
\begin{tabular}{|c|c|c|}
\hline
\textbf{Metric}                    & \textbf{SS} & \textbf{DMC} \\ \hline
\(q^{\text{p,max}}\) [kW]            & 56.0                            & 56.2                              \\ \hline
Estimator COV      & 1\%                             & 1\%                               \\ \hline
Function evaluations                & 2800                            & 9000                              \\ \hline
Solution time [s]                   & 143                             & 435                               \\ \hline
\end{tabular}
\label{table:quantile_estimation_comparison}
\end{table}

\begin{figure}
    \centering
    \includegraphics[width=\linewidth]{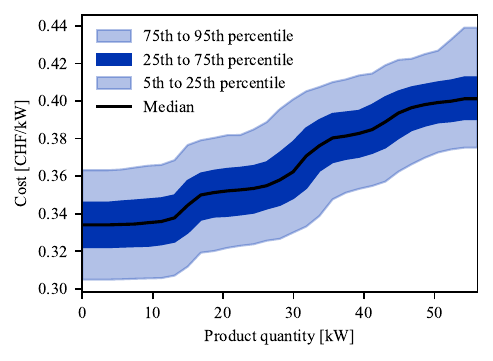}
    \caption{Supply curve of the power reserve capacity product provided by the virtual power plant.}
    \label{fig:supply_curve}
\end{figure}

Fig.~\ref{fig:supply_curve} shows the product supply curve resulting from the flexibility cost assessment model. The curve is presented using a fan chart to highlight the effect of uncertainty on the cost. 
The shape of the curve is monotonically non-decreasing, with average costs ranging from \(\SI{0.34}{CHF/kW}\) to \(\SI{0.40}{CHF/kW}\). The main drivers of the product cost are the opportunity costs arising from the missed participation in the energy markets. 

These opportunity costs stem from the need to guarantee the availability of reserves in real time. To achieve this, the VPP must adopt an energy market schedule that deviates from the cost-optimal one, leading to lost revenue. We observe that the explicit costs of DERs have a second-order effect on reserve capacity costs. This is reflected in the shape of the supply curve by the four distinct cost steps, each corresponding to four different values of the electricity price during the inspected time horizon. When reserve capacity demand is low, the VPP manager redispatches resources only during the time step where the price is the lowest, minimizing opportunity costs. As reserve capacity demand increases, it becomes necessary to redispatch resources during increasingly expensive hours, resulting in the four-step shape in Fig.~\ref{fig:supply_curve}.

\subsection{Impact of the product's requirements}
\label{sec: impact_of_products_requirements}

\begin{figure*}
    \centering
    \captionsetup[subfloat]{position=top}
    \subfloat[]{\includegraphics[width = .25\textwidth,valign=b]{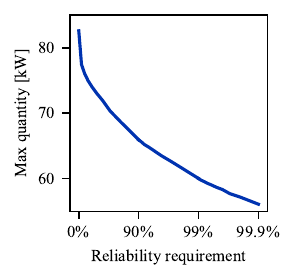}\label{fig:flexibility_reliability_plot}
    \vphantom{\includegraphics[width = .217\textwidth]{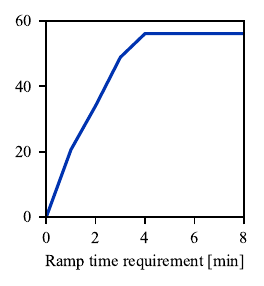}}} 
    \subfloat[]{\includegraphics[width = .22\textwidth,valign=t]{quantity_vs_ramp_rate.pdf}\label{fig:quantity_vs_ramp_rate}}
    \subfloat[]{\includegraphics[width = .225\textwidth,valign=t]{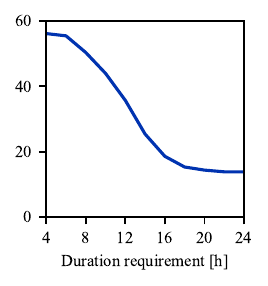}\label{fig:quantity_vs_product_duration}
    \vphantom{\includegraphics[width=0.24\textwidth,valign=t]{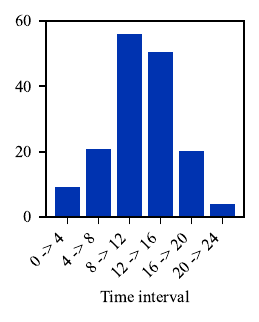}}} 
    \subfloat[]{\includegraphics[width = .22\textwidth,valign=t]{quantity_bar_chart_per_time_interval.pdf}\label{fig:quantity_bar_chart_per_time_interval}}
    \caption{Impact of the product's technical requirements on the maximum power reserve product quantity the virtual power plant can provide.}
    \label{fig:contributions_of_technologies}
\end{figure*}

The following discusses the sensitivity of the maximum offerable product quantity to four technical parameters of the reserve products: reliability, ramp time, duration, and delivery time. 

First, Fig.~\ref{fig:flexibility_reliability_plot} shows the relationship between the maximum available quantity and the reliability requirement of the reserve product. When the reliability requirement is relaxed, the maximum available quantity increases, highlighting the trade-off between the reserve's reliability and the VPP's ability to offer it.

Second, Fig.~\ref{fig:quantity_vs_ramp_rate} shows the impact of the product's ramp time requirement on the maximum available quantity. The highest value is achieved for ramp times exceeding four minutes. For shorter ramps, the PV generation becomes the limiting factor due to the ramping constraints.
These constraints are not related to the technological limitations of PV inverters but arise from LV network connection rules~\cite{verband2020laender}, designed to ensure system stability. These ramping limitations could be relaxed if the PV units are used to provide ancillary services, allowing the VPP to deliver its full reserve capacity with faster ramp times.

Third, Fig.~\ref{fig:quantity_vs_product_duration} illustrates the dependency of the maximum available quantity on the product's duration, assuming the product starts at 8:00 am. The longer the product duration, the lower the maximum available quantity. This reduction occurs because longer durations require reserve provision during low or zero PV production periods. In such cases, the reserve must be provided by other technologies such as HPs, BESS, or EVs, which shift reserve from periods of high PV generation (daylight hours) to those of low PV generation (nighttime).

Finally, Fig.~\ref{fig:quantity_bar_chart_per_time_interval} shows how the maximum product quantity varies across different time intervals during the day for a four-hour reserve product. The quantity is maximum during daylight hours when PV production is high, and curtailment can be used to provide reserve. During nighttime, the availability is significantly lower. Interestingly, the reserve capacity for the 4-hour product at night is lower than that of the 24-hour product shown in Fig.~\ref{fig:quantity_vs_product_duration}. This suggests that selecting longer-duration products can enhance nighttime reserve availability at the expense of reserve availability during the day, highlighting the trade-off between reserve availability during daytime and nighttime hours based on product duration. 

In summary, the technical parameters of the reserve product play a crucial role in determining the maximum available quantity. The model effectively captures these dependencies, suggesting its potential application in the design of DER-specific ancillary services products.

\section{Conclusions}
\label{sec:conclusions}

This paper introduces a novel method to characterize the supply curve of reserve capacity products by technical virtual power plants (VPP) under uncertainty. A two-step approach is proposed that considers network, distributed energy resources (DERs), and product technical constraints. First, the maximum product quantity is characterized, and a novel extreme quantile estimation technique based on subset simulation (SS) is presented, allowing for efficient uncertainty quantification. Second, the supply curve is built through the \(\epsilon\)-constrained approach, considering both explicit and opportunity costs.

The proposed method is demonstrated on a VPP based on a representative Swiss low-voltage network and market setup. The results show that the VPP can reliably provide reserve capacity products. We find that the product's supply curve is monotonically non-decreasing, with opportunity costs from participating in the energy markets being the main cost driver. In addition, we show that the maximum product quantity the VPP can offer strongly depends on the technical requirements of the product. Specifically, we find that strict reliability requirements reduce the available reserve capacity, while longer product durations shift reserve availability from daytime to nighttime. Lastly, we show that the proposed SS-based quantile estimator significantly improves computational efficiency, reducing the burden by \SI{69}{\%} for the analyzed system.

In future work, the model could be applied to a larger pool of networks and products.
Moreover, the proposed SS-based quantile estimation approach could be applied to other uncertainty quantification problems.
Lastly, the proposed method provides a valuable foundation to support VPP managers in designing bidding strategies and policymakers in designing DER-focused AS products, ultimately enabling the integration of higher shares of renewables and thereby supporting the energy transition.

\section*{Acknowledgements}
\label{Acknowledgements}
The author would like to thank Dr. Raphael Wu from Swissgrid AG for the support and insightful discussions. This project was carried out with the support of the Swiss Federal Office of Energy as part of the SWEET EDGE project (https://www.sweet-edge.ch).

\bibliographystyle{IEEEtran}
\bibliography{refs}

\begin{thebibliography}{10}
\providecommand{\url}[1]{#1}
\csname url@samestyle\endcsname
\providecommand{\newblock}{\relax}
\providecommand{\bibinfo}[2]{#2}
\providecommand{\BIBentrySTDinterwordspacing}{\spaceskip=0pt\relax}
\providecommand{\BIBentryALTinterwordstretchfactor}{4}
\providecommand{\BIBentryALTinterwordspacing}{\spaceskip=\fontdimen2\font plus
\BIBentryALTinterwordstretchfactor\fontdimen3\font minus \fontdimen4\font\relax}
\providecommand{\BIBforeignlanguage}[2]{{%
\expandafter\ifx\csname l@#1\endcsname\relax
\typeout{** WARNING: IEEEtran.bst: No hyphenation pattern has been}%
\typeout{** loaded for the language `#1'. Using the pattern for}%
\typeout{** the default language instead.}%
\else
\language=\csname l@#1\endcsname
\fi
#2}}
\providecommand{\BIBdecl}{\relax}
\BIBdecl

\bibitem{xie2024}
Y.~Xie, Y.~Zhang, W.-J. Lee, Z.~Lin, and Y.~A. Shamash, ``Virtual power plants for grid resilience: A concise overview of research and applications,'' \emph{IEEE/CAA Journal of Automatica Sinica}, vol.~11, no.~2, pp. 329--343, Feb 2024.

\bibitem{pudjianto2007}
D.~Pudjianto, C.~Ramsay, and G.~Strbac, ``Virtual power plant and system integration of distributed energy resources,'' \emph{IET Renewable Power Generation}, vol.~1, pp. 10--16, 2007.

\bibitem{camal2023}
S.~Camal, A.~Michiorri, and G.~Kariniotakis, ``Reliable provision of ancillary services from aggregated variable renewable energy sources through forecasting of extreme quantiles,'' \emph{IEEE Transactions on Power Systems}, vol.~38, no.~4, pp. 3070--3084, 2023.

\bibitem{silva2018}
J.~Silva, J.~Sumaili, R.~J. Bessa, L.~Seca, M.~A. Matos, V.~Miranda, M.~Caujolle, B.~Goncer, and M.~Sebastian-Viana, ``Estimating the active and reactive power flexibility area at the tso-dso interface,'' \emph{IEEE Transactions on Power Systems}, vol.~33, no.~5, pp. 4741--4750, 2018.

\bibitem{wang2024}
S.~Wang, W.~Wu, Q.~Chen, J.~Yu, and P.~Wang, ``Stochastic flexibility evaluation for virtual power plants by aggregating distributed energy resources,'' \emph{CSEE Journal of Power and Energy Systems}, vol.~10, no.~3, pp. 988--999, May 2024.

\bibitem{tan2020}
Z.~Tan, H.~Zhong, Q.~Xia, C.~Kang, X.~S. Wang, and H.~Tang, ``Estimating the robust p-q capability of a technical virtual power plant under uncertainties,'' \emph{IEEE Transactions on Power Systems}, vol.~35, no.~6, pp. 4285--4296, 2020.

\bibitem{churkin2024}
A.~Churkin, W.~Kong, J.~N. Melchor~Gutierrez, E.~A. Martínez~Ceseña, and P.~Mancarella, ``Tracing, ranking and valuation of aggregated der flexibility in active distribution networks,'' \emph{IEEE Transactions on Smart Grid}, vol.~15, no.~2, pp. 1694--1711, 2024.

\bibitem{Contreras2021}
D.~A. Contreras and K.~Rudion, ``Computing the feasible operating region of active distribution networks: Comparison and validation of random sampling and optimal power flow based methods,'' \emph{IET Generation, Transmission \& Distribution}, vol.~15, no.~10, pp. 1600--1612, 2021.

\bibitem{Riaz2022}
S.~Riaz and P.~Mancarella, ``Modelling and characterisation of flexibility from distributed energy resources,'' \emph{IEEE Transactions on Power Systems}, vol.~37, no.~1, pp. 38--50, 2022.

\bibitem{gonzales2018}
D.~Mayorga~Gonzalez, J.~Hachenberger, J.~Hinker, F.~Rewald, U.~Häger, C.~Rehtanz, and J.~Myrzik, ``Determination of the time-dependent flexibility of active distribution networks to control their tso-dso interconnection power flow,'' in \emph{2018 Power Systems Computation Conference (PSCC)}, 2018, pp. 1--8.

\bibitem{kalantarneyestanaki2020}
M.~Kalantar-Neyestanaki, F.~Sossan, M.~Bozorg, and R.~Cherkaoui, ``Characterizing the reserve provision capability area of active distribution networks: A linear robust optimization method,'' \emph{IEEE Transactions on Smart Grid}, vol.~11, no.~3, pp. 2464--2475, May 2020.

\bibitem{maMmaRELLA2022110108}
M.~Mammarella, V.~Mirasierra, M.~Lorenzen, T.~Alamo, and F.~Dabbene, ``Chance-constrained sets approximation: A probabilistic scaling approach,'' \emph{Automatica}, vol. 137, p. 110108, 2022.

\bibitem{zhou2024}
Y.~Zhou, C.~Essayeh, and T.~Morstyn, ``Aggregated feasible active power region for distributed energy resources with a distributionally robust joint probabilistic guarantee,'' \emph{IEEE Transactions on Power Systems}, pp. 1--15, 2024.

\bibitem{swissgrid2022_ASproducts}
\BIBentryALTinterwordspacing
{Swiss Grid}, ``Principles of ancillary services products,'' {Swiss Grid}, Tech. Rep., 2022, accessed: 2024-10-23, Version 19, 16 December 2022. [Online]. Available: \url{https://www.swissgrid.ch/content/dam/swissgrid/customers/topics/ancillary-services/as-documents/D220824-AS-Products-V19-en.pdf}
\BIBentrySTDinterwordspacing

\bibitem{swissgrid2022_prequalification}
\BIBentryALTinterwordspacing
------, ``Anhang: Praequalifikationsbedingungen für die teilnahme an der primär-, sekundär- und tertiärregelung,'' {Swiss Grid}, Tech. Rep., 2022, accessed: 2024-10-23. [Online]. Available: \url{https://www.swissgrid.ch/content/dam/swissgrid/customers/topics/ancillary-services/prequalification/2/Anhang-01-Praequalifikationsbedingungen-de.pdf}
\BIBentrySTDinterwordspacing

\bibitem{guyader2011simulation}
A.~Guyader, N.~Hengartner, and E.~Matzner-Løber, ``Simulation and estimation of extreme quantiles and extreme probabilities,'' \emph{Applied Mathematics \& Optimization}, vol.~64, no.~2, pp. 171--196, 2011.

\bibitem{zuev2013subset}
K.~M. Zuev, ``Subset simulation method for rare event estimation: An introduction,'' \emph{Encyclopedia of Earthquake Engineering}, 2013.

\bibitem{doi:https://doi.org/10.1002/9781118398050.ch5}
S.-K. Au and Y.~Wang, \emph{Subset Simulation}.\hskip 1em plus 0.5em minus 0.4em\relax John Wiley \& Sons, Ltd, 2014, ch.~5, pp. 157--204.

\bibitem{19266}
M.~Baran and F.~Wu, ``Optimal sizing of capacitors placed on a radial distribution system,'' \emph{IEEE Transactions on Power Delivery}, vol.~4, no.~1, pp. 735--743, 1989.

\bibitem{amara2015comparison}
F.~Amara, K.~Agbossou, A.~Cardenas, Y.~Dubé, and S.~Kelouwani, ``Comparison and simulation of building thermal models for effective energy management,'' \emph{Smart Grid and Renewable Energy}, vol.~6, no.~4, pp. 95--113, 2015.

\bibitem{6099519}
X.~Fang, S.~Misra, G.~Xue, and D.~Yang, ``Smart grid — the new and improved power grid: A survey,'' \emph{IEEE Communications Surveys \& Tutorials}, vol.~14, no.~4, pp. 944--980, 2012.

\bibitem{dong_quantile_2017}
H.~Dong and M.~K. Nakayama, ``Quantile estimation with latin hypercube sampling,'' \emph{Operations Research}, vol.~65, no.~6, pp. 1678--1695, 2017.

\bibitem{4308298}
Y.~Y. Haimes, L.~S. Lasdon, and D.~A. Wismer, ``On a bicriterion formulation of the problems of integrated system identification and system optimization,'' \emph{IEEE Transactions on Systems, Man, and Cybernetics}, vol. SMC-1, no.~3, pp. 296--297, 1971.

\bibitem{7530870}
M.~Parvania and R.~Khatami, ``Continuous-time marginal pricing of electricity,'' \emph{IEEE Transactions on Power Systems}, vol.~32, no.~3, pp. 1960--1969, 2017.

\bibitem{GORMAN2022105832}
W.~Gorman, C.~C. Montañés, A.~Mills, J.~H. Kim, D.~Millstein, and R.~Wiser, ``Are coupled renewable-battery power plants more valuable than independently sited installations?'' \emph{Energy Economics}, vol. 107, p. 105832, 2022.

\bibitem{He2018}
G.~He, Q.~Chen, P.~Moutis, S.~Kar, and J.~F. Whitacre, ``An intertemporal decision framework for electrochemical energy storage management,'' \emph{Nature Energy}, vol.~3, no.~5, pp. 404--412, 2018.

\bibitem{ACER2023}
{Agency for the Cooperation of Energy Regulators}, ``Report on electricity transmission and distribution tariff methodologies in europe,'' ACER, Tech. Rep., January 2023.

\bibitem{Oneto_2023}
A.~E. Oneto, B.~Gjorgiev, F.~Tettamanti, and G.~Sansavini, ``Large-scale inference of geo-referenced power distribution grids using open data,'' \emph{TechRxiv}, Dec. 2023.

\bibitem{EnergyPerspectives2050+}
\BIBentryALTinterwordspacing
{Swiss Federal Office of Energy (SFOE)}, ``Energy perspectives 2050+,'' SFOE, Tech. Rep., December 2021, accessed: 23/10/2024. [Online]. Available: \url{https://www.bfe.admin.ch/bfe/en/home/policy/energy-perspectives-2050-plus.html}
\BIBentrySTDinterwordspacing

\bibitem{entsoe_transparency}
ENTSO-E, ``Transparency platform,'' \url{https://newtransparency.entsoe.eu/market/energyPrices}, 2024, accessed: 2024-10-23.

\bibitem{ewz_2024_tariff}
EWZ, ``Stromtarife 2024,'' \url{https://ewz.ch/dam/ewz/Privatkunden/Strom/Tarife/Dokumente/Uebersicht_Tarife_ewz_2024.pdf}, Tramstrasse 35, 8050 Zürich, 2024, accessed: 2024-10-23.

\bibitem{6520086}
B.-M. Hodge, D.~Lew, and M.~Milligan, ``Short-term load forecast error distributions and implications for renewable integration studies,'' in \emph{2013 IEEE Green Technologies Conference (GreenTech)}, 2013, pp. 435--442.

\bibitem{VANDERMEER20181484}
D.~{van der Meer}, J.~Widén, and J.~Munkhammar, ``Review on probabilistic forecasting of photovoltaic power production and electricity consumption,'' \emph{Renewable and Sustainable Energy Reviews}, vol.~81, pp. 1484--1512, 2018.

\bibitem{10694497}
M.~P. Herrera, M.~Schwarz, and G.~Hug, ``Spatio-temporal modeling of large-scale bev fleets' charging energy needs and flexibility,'' in \emph{2024 International Conference on Smart Energy Systems and Technologies (SEST)}, 2024, pp. 1--6.

\bibitem{8039509}
W.~Kong, Z.~Y. Dong, Y.~Jia, D.~J. Hill, Y.~Xu, and Y.~Zhang, ``Short-term residential load forecasting based on lstm recurrent neural network,'' \emph{IEEE Transactions on Smart Grid}, vol.~10, no.~1, pp. 841--851, 2019.

\bibitem{solcast_forecast_accuracy}
Solcast, ``Live and forecast accuracy: Bias and error validation of solcast data,'' \url{https://www.solcast.com/forecast-accuracy}, Solcast, 2024, accessed: 2024-10-23.

\bibitem{meteoblue_weather_accuracy}
Meteoblue, ``Weather data accuracy,'' Available at: \url{https://content.meteoblue.com/en/research-education/weather-data-accuracy}, Meteoblue, 2024, accessed: 2024-10-23.

\bibitem{LAGO2021116983}
J.~Lago, G.~Marcjasz, B.~{De Schutter}, and R.~Weron, ``Forecasting day-ahead electricity prices: A review of state-of-the-art algorithms, best practices and an open-access benchmark,'' \emph{Applied Energy}, vol. 293, p. 116983, 2021.

\bibitem{hart2011pyomo}
W.~E. Hart, J.-P. Watson, and D.~L. Woodruff, ``Pyomo: modeling and solving mathematical programs in python,'' \emph{Mathematical Programming Computation}, vol.~3, no.~3, pp. 219--260, 2011.

\bibitem{gurobi}
\BIBentryALTinterwordspacing
{Gurobi Optimization}, ``{Gurobi Optimizer Reference Manual},'' 2024, accessed: 2024-10-23. [Online]. Available: \url{https://www.gurobi.com}
\BIBentrySTDinterwordspacing

\bibitem{10.5555/898758}
M.~P. Forum, ``Mpi: A message-passing interface standard,'' University of Tennessee, USA, Tech. Rep., 1994.

\bibitem{verband2020laender}
\BIBentryALTinterwordspacing
V.~S. Elektrizitätsunternehmen, ``Ländereinstellungen schweiz 2020,'' 2020, accessed: 2024-10-23. [Online]. Available: \url{https://www.strom.ch/de/media/12432/download}
\BIBentrySTDinterwordspacing

\end{thebibliography}

\end{document}